\documentclass[twocolumn,12pt,cites,graphicx]{article}
\usepackage{epsfig}
\usepackage{bm,bbm,amssymb,color, amsmath, color}

\newcommand{\vect}[1]{\boldsymbol{#1}}

\newcommand{\dif}[1]{\textrm{d} #1 \ }

\newcommand{\eq}[1]{\begin{eqnarray}
{#1}
\end{eqnarray}}

\title{Buckling of semiflexible filaments under compression}
\author{J. R. Blundell and E.M. Terentjev \\ 
{\small Cavendish Laboratory, University of Cambridge} \\
{\small JJ Thomson Avenue, Cambridge, CB3 0HE, U.K.} }

\begin{document}

\maketitle
\renewcommand{\thefootnote}{\fnsymbol{footnote}}

\noindent A model for filament buckling at finite temperatures is presented. {Starting from the classical} worm-like chain model under constant compression, we use a mean-field approach for filament inextensibility to find the complete partition function. We find that there is a simple interpolation formula that describes the free energy of chains or filaments as a function of end-to-end separation, {which spans the whole range of filament stiffnesses}. Using this formula we study the buckling transition of semiflexible filaments and find that kinetics plays an important role. We propose that the {filament buckling is essentially the first order transition} governed by the kinetics of escaping a local free energy minimum. A simple model for the kinetics is put forward, {which} shows the critical buckling force for a filament is reduced by a fraction that has a universal scaling with temperature with an exponent $\nu = 0.56$.

\section{Introduction}

There {are a number of reasons for the current interest in semiflexible filaments under compression}. Firstly the three structural components of the cytoskeleton: microtubules, {intermediate and} actin filaments are all classed as semiflexible filaments. Experiments on in-vitro microtubules \cite{Ingber, Kurachi} and recent experiments on in-vitro dendritic actin \cite{Fletcher}, as well as experiments on in-vivo actin filaments \cite{Costa}, have shown that filaments {buckling} under compression often have a crucial role in determining the behavior of the cellular network. In addition there is considerable interest in the mechanical properties of carbon nanotubes (CNT's) under compression \cite{Falvo, Yap}, with views to application in drug delivery \cite{Kostarelos1}. In light of  {these considerations}, understanding the behavior of such filaments when subject to compressive forces is of considerable importance.

The buckling of rods under longitudinal compression is a classic problem dating back to Euler \cite{euler}. For a macroscopic {elastic} rod with bending modulus $A$ and length $L$ that is pinned at both ends, Euler derived that above a critical force $f_c = A \pi^2 / L^2$ the rod can no longer support the compressive force and catastrophically buckles \cite{Landau, Feynman}. {The Euler buckling threshold only depends on the bending modulus of the rod, the compressive (Young) modulus does not enter directly into the expression. In fact, a simple analysis suggests that the filament extensibility can be safely ignored, for homogeneous filaments or hollow tubes alike. Consider an elastic tube of outer radius $r$ and the wall thickness $b$ (the homogeneous solid rod would have $b=r$) made of a material with the Young modulus $Y$. The bending modulus of such a filament is $A \sim \int Y r^3 \dif{r} \sim (r^3 b/L)Y$ \cite{Landau} and the characteristic energy scale of small bending is $A/L$, see Figure~\ref{sketch}. The equivalent strain can be achieved by longitudinal compression, with a characteristic energy scale $\sim (rbL)Y$. The ratio of bending to compression energies is, therefore, $(r/L)^2$. For most semiflexible filaments, radii are on the order of $nm$ while lengths are on the order of microns, so that $(r / L)^2 \sim 10^{-6}$, therefore, most of the imposed deformation will be accommodated in bending, so the compressibility of filaments can be neglected. We carry on here with the simpler problem of incompressible filaments.}

If we reduce the scale of rods such that we are now dealing with microscopic filaments, there comes a regime where the characteristic bending energy of the filament $A / L$ is of the same order as the thermal energy  {supplied by} the environment, $k_B T$. Under such conditions we can no longer only consider the {mechanical} energy of the system, {but} {must} also include the effect of entropy. This is the main purpose of the current work.

The effect of temperature on the buckling transition has been considered before, notably in the works \cite{Odijk, Padgornik, Lipowsky, Emanuel}. Odijk evaluates the partition sum as a semi-classical {series} expansion in three dimensions while Baczynski et. al. \cite{Lipowsky} and Emanuel et. al. \cite{Emanuel} consider contributions to the partition sum from anharmonic terms in two and three dimensions respectively. Hansen et. al. \cite{Padgornik} evaluate the partition sum by means of a saddle point approximation controlled by the small parameter $1/d$ where $d$ is the spatial dimension and focus largely on the effect of non-local interactions along the chain.

In the present work we take a different approach. Invoking a mean-field approximation to filament inextensibility, we calculate the complete partition function and find a simple interpolation formula for the free energy of single filaments under compression that delineates clearly the role of filament bending {rigidity} and the role of entropy, {or local thermal motion}.
For arbitrarily small compressive forces we find that the free energy function possesses two minima,  {one corresponding to the original extended filament, somewhat compressed by the applied force, while the other describing the completely buckled filament. We then propose a simple model that describes the first order transition from unbuckled to buckled states of the {fluctuating} filament. This simple model predicts that the critical buckling force is reduced at non-zero temperatures  {from the original Euler expression}. Our results bear resemblances to earlier work \cite{Emanuel} in predicting a reduction the the critical buckling force at non-zero temperatures, but our mechanism is different: kinetics {of overcoming {a} free energy barrier} plays a crucial role.

The paper is organized in the following way. In subsection \ref{WLC} we {summarize} the well explored worm-like chain model, {which we use with} a global inextensibility approximation to calculate the  {full} partition function. We find a simple interpolation formula that captures all relevant features of  {this partition function, and the corresponding} free energy of single chains under compression in a closed analytical expression valid across the whole range, from flexible Gaussian chains to completely stiff elastic rods; this is shown in subsection \ref{mean_field}. Using this formula we examine the effect of compression on the separation of the ends of the chain in subsection \ref{compression} and develop a simple model for the kinetics of the transition in subsection \ref{kinetics}. We illustrate {and discuss} the main results of this model in section \ref{results}.

\begin{figure}
\centering
\resizebox{0.4\textwidth}{!}{\includegraphics{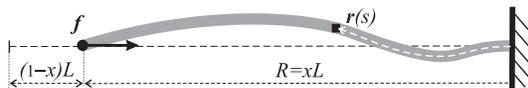}}  \caption{Sketch of a semiflexible filament under compression force $f$, illustrating the notations used in the model.}  \label{sketch}
\end{figure}


\section{Theory} \label{theory}

\subsection{Semiflexible filaments} \label{WLC}

The semiflexible nature of filaments is a result of there being an energy penalty associated with filament curvature. This energy penalty is controlled by the bending modulus $A$ with units of energy $\times$ length.
The {characteristic} bending energy of a filament bending {over length scales of order} $L$ is therefore $A / L$.
For a large class of microscopic filaments, this energy scale is of order $k_B T$, and therefore thermal fluctuations cause the filament to bend over length scales {comparable with} $L$. Such filaments are termed semiflexible. In this regime thermal fluctuations are large enough to weakly bend the filament and thus the filament occupies a middle ground between the flexible Gaussian chains (with no {or weak} bending energy) and rigid {elastic} rods whose bending energy far exceeds $k_B T$.

The statistical physics of such semiflexible filaments can be captured by the so called worm-like chain model first proposed by Kratky and Porod \cite{Kratky} and subsequently developed in further works \cite{Odijk, Fixman, Kleinert, Ha, Frey, Blundell}. In this model, the filament is treated as a space-curve $\vect{r}(s)$, where $s$ parameterizes the contour length  {along the curve}. The Hamiltonian is obtained by associating an energy that scales like the squared local curvature of the space-curve  {and is proportional to the bending modulus of the filament $A$ which measures resistance to bending}:
\eq{H[\vect{r}(s)] = \int_0^L \dif{s} \frac{A}{2} \left( \frac{{\rm d}^2\vect{r}(s)}{{\rm d}{s}^2} \right)^2 \, .}
Here the second derivative with respect to {the arc length} $s$ represents the curvature, and we also have the constraint that the tangent vector has unit length, $({\rm d}{\vect{r}}/{\rm d}{s})^2 = 1$ for all $s$, which ensures that the filament is locally inextensible.

In most situations we are interested in the  {constrained partition function, which is equivalent to the} probability that the ends of the filament are separated by a span vector $\vect{R}  {=\vect{r}(L) - \vect{r}(0)}$ given a filament length $L$. We use $\vect{x}$ to denote the dimensionless measure of end-to-end separation,  $\vect{x} = \vect{R} / L$. In the remainder of the paper we will be interested in the quantity $P(\vect{x}; L)$,  {which is} the probability of a filament of length $L$ adopting a configuration that has its ends separated by $\vect{x}$. This can be obtained by a functional integral over the field $\vect{r}(s)$:
\eq{P(\vect{x};L) \propto \int \mathcal{D} \vect{r} \ \textrm{exp}(-H[\vect{r}] / k_B T) \, , \label{full}}
subject to the  {explicit} constraints:
\eq{ \frac{\vect{r}(L) - \vect{r}(0)}{ L} = \vect{x}   \qquad \textrm{and} \qquad  \left( \frac{{\rm d}{\vect{r}}}{{\rm d}{s}} \right)^2 = 1 \, .}

Physically we are summing over the probabilities of all configurations of the space-curve that start at $0$ and end at $\vect{x}$, and which are locally inextensible. The full expression (\ref{full}) corresponds to a non-linear sigma model, which can only be solved as a series expansion in the small parameter $k_B T L / A  { \ \ll 1}$, and is therefore only valid for very stiff filaments \cite{Frey, Kleinert2}.

\subsection{Mean field inextensibility} \label{mean_field}

In this paper we relax the constraint of  {rigid} local inextensibility to one of global inextensibility that only counts contributions to the functional integral whose \textit{average} tangent vectors are of unit length; $\langle |{\rm d}{\vect{r}}/{\rm d}{s} | \rangle = 1$. This method amounts to a {commonly used} mean-field approach to the problem, as outlined in \cite{Ha, Blundell}. This approach is useful because it makes the problem mathematically tractable and produces closed forms for the probability distribution,  {while retaining the relevant physical features of the filament}. This procedure is analogous to the transformation from microcanonical to canonical ensemble in statistical mechanics. Using this approximation the expression for the probability distribution reduces to
\eq{P(\vect{x};L) \propto \int \mathcal{D} \vect{r} \  \textrm{e}^{-H[\vect{r}] / k_B T} \delta[\langle ({\rm d}{\vect{r}}/{\rm d}{s})^2 \rangle - 1] \, , \label{mean-field} }
subject to the boundary conditions $(\vect{r}(L) - \vect{r}(0)) / L = \vect{x} $. This functional integral can be performed with no further approximation and the result can be written as an integral over an auxiliary field $\phi$ \cite{Blundell}. Moreover, the probability distribution  {turns out to depend} on its variables {only} through a single  {non-dimensional combination of parameters:
$$\gamma = \frac{A}{2k_B T L}(1- x^2).$$}
For a chain in $d$ dimensions the {general} result is:
 \eq{P(\gamma) = \int \dif{\phi}\, e^{i \gamma \phi}{ \left(\frac{\sqrt{i \phi}}{\sin \sqrt{i \phi}}\right)^{d/2}} \label{mean-field}.}
This expression  {(for the non-normalized probability)} uses only the approximation of global inextensibility. It is valid for all {values of bending stiffness} and all extensions of the filament. In two dimensions the integral can be solved analytically via contour integration \cite{Blundell}, and closely agrees with previous expressions obtained in the literature \cite{Frey}. In three dimensions the integral cannot be performed analytically, however, {is easily evaluated numerically due to its dependence only on a single parameter $\gamma$. In the past there have been several approximate analytical expressions used to represent the probability $P(\vect{x};L)$ in 3-dimensions, some used quite widely \cite{Marko,Ha}. After a new analysis} we find that the simple interpolation of the form
\eq{P(\gamma) = \exp[-\pi^2 \gamma - \frac{1}{\pi \gamma}] \label{interpolation}}
is a very good approximation to the {numerical} integral (\ref{mean-field}) for $d=3$. This form captures all essential features of the curve, scaling in the correct way for large and small {values of} $\gamma$. A comparison of the interpolation formula (\ref{interpolation}) with the numerically {evaluated} values of the integral (\ref{mean-field}) in 3-d is shown in Figure~\ref{interpolationplot}. For comparison, this figure also plots the famous and widely used formulas due to Marko and Siggia \cite{Marko}, and Ha and Thirumalai \cite{Ha}. All these, and other analytical expressions agree in the limit $\gamma \rightarrow 0$, which corresponds to the long or flexible chains, eventually reaching the Gaussian limit. There are, however, spectacular disagreements in the limit of stiff, or short chains (at $\gamma \geq 1$), where the limit of pure bending elasticity is not present.

\begin{figure}
\centering
\resizebox{0.4\textwidth}{!}{\includegraphics{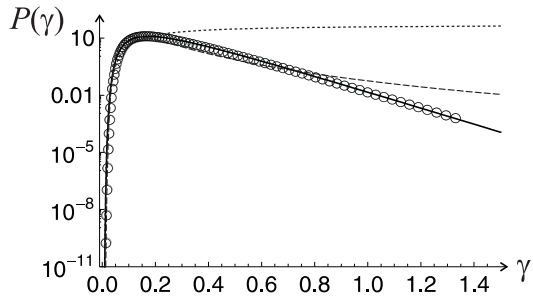}}  \caption{Comparison of numerical values for the 3-d integral in (\ref{mean-field}) (empty circles)  with the interpolation formula (\ref{interpolation}), plotted as a {solid} line. The interpolation fits very well, {which is emphasized by the logarithmic axis}, in particular capturing the correct scaling at large and small $\gamma$ with the maximum at the correct value.  For comparison, dashed line shows the Ha-Thirumalai expression \cite{Ha} and the dotted line the Marko-Siggia expression \cite{Marko}.} \label{interpolationplot}
\end{figure}

Substituting in the expression for $\gamma$, we find that within the global inextensibility approximation there is a simple expression for the free energy {$F=-k_BT \ln P(x)$} of a single chain as a function of extension, given by
\eq{F(x) = \frac{A \pi^2}{2 L}(1-x^2) + \frac{2(k_B T)^2 L}{\pi A (1-x^2)} \, . \label{free1}}

\begin{figure}
\centering
\resizebox{0.4\textwidth}{!}{\includegraphics{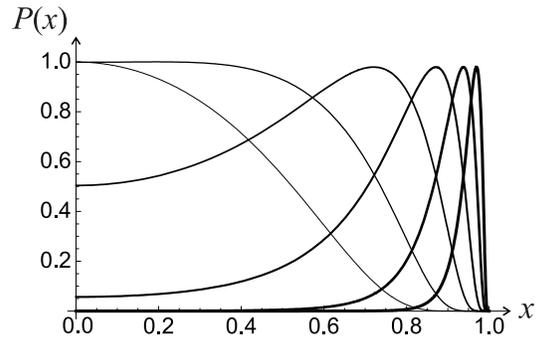}}  \caption{Non-normalized probability distribution plots $P(x)$ for filaments of increasing stiffness (in order of increasing line width): $A/k_B T L = 0.125, 0.25, 0.5, 1.0, 2.0, 4.0$. } \label{probplots}
\end{figure}

This {form of the chain free energy} differs from previous results obtained in the literature \cite{Marko,Ha,Kleinert} in that it clearly delineates the role of bending energy and the non-zero temperature effects,  {apart from matching the numerical result almost exactly (in fact, we are somewhat mystified by the perfect matching of the fit given by the expression (\ref{interpolation}), which suggest that there is an underlying analytical route to this integration)}. As one would expect  {in the retrospect, the free energy (\ref{free1}) has} a term independent of temperature that is simply the internal energy of the bent filament, and  {also} a term that captures the {effect of thermal fluctuations}. We note that this is not simply an entropic term since it is proportional to $T^2$, it contains a mixture of energetic effects and entropic effects at non-zero temperature.

If one takes the limit of a very short filament, or equivalently a filament with a {very} large bending modulus $A$, then the energetic term dominates over the entropic term, and the free energy of the filament is simply the bending energy of {an elastic} rod. In this limit the formula (\ref{free1}) reassuringly recovers the bending energy of a macroscopic rod that is derived in the small bending regime using a variational principle: $U \approx (A \pi^2 /L)(1-x)$ \cite{Timoshenko}.

In the {opposite} limit of small bending modulus or a very long filament, the entropic term dominates the free energy. In this regime the minimum of the free energy is at zero extension and, expanding about this minimum, in the leading order one recovers the free energy of a Gaussian chain with step length (or the persistence length) $l_p \sim A / k_B T$. Unlike a true Gaussian chain however, higher order terms in our expansion would implement the inextensibility of the chain that we have imposed. We illustrate these limits and the crossover between them by plotting the non-normalized probability distributions for filaments of increasing stiffness in Figure~\ref{probplots}.

\subsection{Filament under compression} \label{compression}

From the {compact analytical} expression for the free energy of a single filament (\ref{interpolation}) we are able to write down the free energy for a {situation when this filament is subject to a} constant compressional force $f$. 
 {In this case} there is an additional energetic contribution due to the work done by the compressive force in moving the ends of the filament parallel to the applied force. In order that the filament does not rotate, this force must be directed parallel to the span vector of the filament, and therefore the work done is ${\Delta W=}f L x$, and so the free energy of a single filament {takes the form}
\begin{align}
F(x, f) =  \ & f L x + \frac{A \pi^2}{2L}(1-x^2) +  \frac{2(k_B T)^2 L}{\pi A (1-x^2)}  \label{free_energy}.
\end{align}
where the positive $f$ refers to a compressional force.  Figure~\ref{free_plots} shows the free energy of a semiflexible filament ({at a fixed} $A / k_B T L  = 3$) plotted as a function of the relative separation for increasing compressive forces.

\begin{figure}
\centering
\resizebox{0.4\textwidth}{!}{\includegraphics{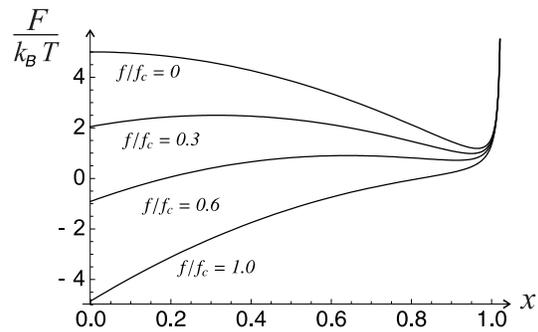}}  \caption{$F(x; f)$ plotted for a typical semiflexible filament ($A / k_B T L = 3$) subject to increasing compressive forces measured in units of the Euler critical force $f_c = A \pi^2/ L^2$. The four regimes of one minimum (unbuckled), two minima (unbuckled lower), two minima (buckled lower) and one minimum (buckled) are illustrated by the four curves (low to high). We stop the plots at $x=0$ whereas in reality they could extend to $x<0$ which corresponds to the ends ``swapping over" and the filament being subject to tension. Such effects are not addressed in this paper.}  \label{free_plots}
\end{figure}

Examining Figure~\ref{free_plots} we see there are four regimes. For no compression $f = 0$, the free energy possesses a single minimum near full extension corresponding to an unbuckled filament.  {This minimum, corresponding to the equilibrium end-to-end distance of a semiflexible filament, is at $x_{\rm eq}=(1- 2k_B T L/ \pi^{3/2} A)^{1/2}$. The linear modulus $k$ (an effective spring constant, that is the curvature of the free energy at $x_{eq}$) of such a filament at small extension/compression forces about its equilibrium is
\eq{k = 4 \pi^{5/2} \frac{k_B T l_p^2 }{L^4} \left(1- \frac{2L}{\pi^{3/2} l_p }\right) \label{linear}}
where $l_p = A / k_B T$ is what one traditionally calls the persistence length in longer or more flexible chains. Clearly, below a certain bending rigidity (or for long enough filaments, $L/l_p \geq \pi^{3/2} / 2$) the equilibrium separation $x_{\rm eq}=0$ and the filament responds to stretching with purely entropic elasticity, as classical polymer chains, see Figure~\ref{probplots}. All of these conclusions are very easy to obtain from the new compact analytical expression for the free energy (\ref{free_energy}).}

{Returning to semiflexible filaments,} for intermediate compressions $ 0 < f \leq f_1(T)$ the free energy has two minima - one corresponding to an unbuckled filament (close to {the full extension} $x=1$) and the other corresponding to a buckled filament (at $x=0$). Provided the compression does not exceed $f_1$, a critical force that depends on temperature, the minimum of the {extended} filament is lower in free energy than the buckled free energy {and so nothing dramatic will happen to it}: {the filament will resist the compression force as an effective spring with a constant given by (\ref{linear}) at $f \rightarrow 0$.}

For $ f \geq f_2(T) $ there is only one free energy minimum at $x=0$ corresponding to the buckled state {and so the filament cannot sustain the applied force}. {In the region} $ f_1(T) < f \leq f_2(T) $ the free energy has two minima, {with} the minimum corresponding to a buckled filament lower in free energy; {the buckling transition may occur anywhere in this region depending on the relative height of the energy barrier separating the {metastable extended state}.
This scenario clearly resembles the first-order phase transition with a discontinuous jump of the ``order parameter'' and a region of hysteresis.} A free energy function that possesses two minima of equal depth, at $f=f_1$, would be characteristic of {the equilibrium point of the} first order phase transition. However, {as in all discontinuous transitions}, the {passage} from an unbuckled configuration to a buckled one means climbing over a free energy barrier. If the system is unable to {pass over} this barrier in timescales of observation, the transition will not occur. The kinetics of this transition will therefore be crucial to what is observed experimentally. The force $f_1$ is not necessarily the force at which buckling will occur, instead we are interested in the force, which will be called $f^*$ where the kinetics of the buckling transition changes from being slow to fast (in comparison to observation times).

\subsection{Transition Kinetics} \label{kinetics}

The transition from the {extended filament} to the buckled one means climbing over a free energy barrier, the height of which depends on the applied compression.  The rate of escape over this barrier is {the classical} Kramers escape problem \cite{Kramers}. The rate of escape $\tau$ from a metastable local minimum {to the true equilibrium state} is dominated by an exponential term in the barrier height $\Delta F$,  {i.e.} $\tau = \tau_0 \exp(-\Delta F / k_B T)$. If the barrier height is far greater than $k_B T$ then the kinetics of the transition will be very slow. Conversely the escape will take place at an appreciable rate if the thermal energy is of a similar magnitude to the barrier, $\Delta F \sim k_B T$. {Strictly, this is a delicate kinetic problem involving a hierarchy of attempt rates $\tau_0$ for different chain segments. However, here we take a simple and qualitative approach, relying on the strong effect of the activation exponential term. Let us denote $f^*$ the value of external force at which $\Delta F(f^*,T) = k_B T$, see Figure~\ref{barrier}.}  At this point we would expect there to be enough thermal energy in the environment to excite the filament out of the local minimum. For compression forces above $f^*$ we would expect the filament to lie in the lower global minimum of the buckled state. {We therefore propose the rate of escape follows an ``all or nothing" transition at $f=f^*$ defined by the conditions}
{
\begin{equation}
\begin{array}{cl}
\Delta F(f, T) > k_B T & \ \ \textrm{no transition}\\
\Delta F(f, T) < k_B T & \ \ \textrm{transition}\\
\end{array}
\end{equation} }
where there is equality when $f=f^*$.

\begin{figure}
\centering
\resizebox{0.4\textwidth}{!}{\includegraphics{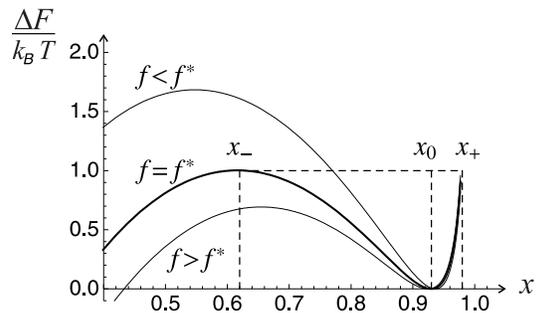}}  \caption{A plot of the free energy of a single semiflexible filament ($A / k_B T L = 3$) subject to a compressive forces about $f=f^*$, {focusing on the region near the extended state}. At $f=f^*$ (thick line) the barrier is exactly $k_B T$ {high}. For compressive forces $f < f^*$ the barrier height is greater than $k_B T$ and the filament in confined to the local minimum in separation between $x_+$ and $x_-$; it remains unbuckled. For forces $f > f^*$ the barrier height is lower than $k_B T$ and {we assume} the thermal energy is enough for the filament to ``escape'' the {metastable state} and buckle.} \label{barrier}
\end{figure}

It should be stressed that this ``all-or-nothing'' model is really an oversimplification. The transition to a buckled state can proceed {even} if $k_B T < \Delta F$ and, {equally,} a filament can remain unbuckled even if $k_B T > \Delta F$, depending on the time of observation (or the rate of force application). However the model does capture the important physics that it is only in the vicinity of $\Delta F = k_B T$ that the rate of the transition changes from very slow to very fast. More precisely, if the rate of escape is of the Kramers' form $\tau \sim \tau_0 \exp(-\Delta F / k_B T)$ then the change from ``slow" to ``fast" kinetics occurs at the inflection point of $\tau (T)$, at $\Delta F = 2 k_B T$; the characteristic width of this transition is $\sim k_B T$. We are justified in assuming the transition rate is a step function at $\Delta F = k_B T$ if the width of the maximum ($\sim k_B T$) is small in comparison to the overall energy scale of the filament $A/L$.

Adopting this simple view of the kinetics therefore places a limit on the range of temperatures for which this model will be {practically} valid. The energy scale that determines the height of the barrier is $A/L$. If we are in a regime where {$A/L < k_B T$} then no matter what compressive force we impose, the filament is able to explore its entire range of extensions in short times. In other words, the flexible chain is able to explore most of its conformations in the time of observation, which is the basis of classical polymer physics. The filament therefore equilibrates rapidly and the assumption that it is trapped in a  {metastable} local minimum is a poor one.
{If the system is even more flexible,  $A/L < 0.5 k_B T$, then it cannot support any compressional force at all.} Our assumption that the transition is a non-equilibrium ``all-or-nothing'' type transition limited by the kinetics of the transition, is only valid if $A/L > k_B T$, that is, with no force applied, the filament must be confined to the unbuckled minimum. For the remainder of the paper therefore, we will focus on regime {of relatively stiff filaments} where $A / k_B T L > 1$.

\section{Results and discussion} \label{results}

\begin{figure}
\centering
\resizebox{0.4\textwidth}{!}{\includegraphics{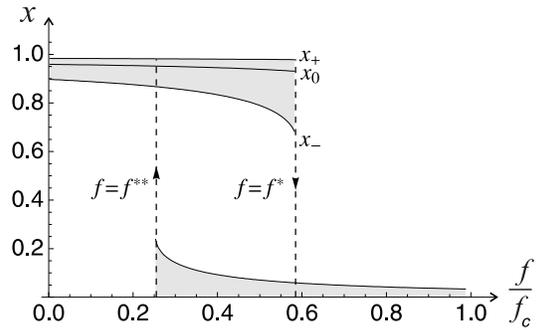}}  \caption{Plots of $x_+$, $x_-$ and $x_0$ for a semiflexible filament ($A/ k_B T L = 3$) subject to increasing compression measured in units of the critical Euler buckling force $f_c$. For increasing compression, at $f=f^*$ there is no longer a solution for $x_-$ and the filament buckles. On decreasing compression the reverse process occurs at a lower force $f=f^{**}$. Expressions for $f^*$ and $f^{**}$ are given in the text.} \label{hysteresis}
\end{figure}

In this section we discuss the main results of {our analysis of the filament} buckling transition. With this ``all or nothing'' model defined, the transition to a buckled state occurs at a critical force $f^*(T)$ defined by $\Delta F (f^*, T) = k_B T$, i.e. the force at which the barrier height is $k_B T$. This is illustrated in Figure~\ref{barrier}. Below $f^*$, the filament is confined to the local minimum centered on $x_0$ of the unbuckled state, and can explore the vicinity of the minimum up to points $x_+$ and $x_-$ defined as the separations at which the free energy is $k_B T$ higher than at the local minimum. As the force $f \rightarrow f^*$, from below, both $x_+$ and $x_-$ decrease (see Figure \ref{hysteresis}), until the point $f=f^*$ when there is no longer a solution for $x_-$ and the filament makes the transition to the buckled state. Overall, the range between $x_+$ and $x_-$ represents the characteristic range of fluctuations of the filament span,  {which is shown as the shaded area in Figure~\ref{hysteresis}.}

Figure~\ref{hysteresis} shows a plot of $x_+$, $x_-$ and $x_0$  {for the applied force increasing} up until the buckling transition at $f=f^*$. Above this force the filament {catastrophically} buckles to a separation $x=0$ and can once again explore the free energy minimum up to points that are $k_B T$ higher in energy than the minimum. If we reverse this and now start to reduce the {originally applied} compression force, there comes a point $f=f'$ where the minimum at $x \approx 1$ (unbuckled) becomes lower in free energy than the buckled one at $x=0$. However, just as for the case of buckling, there is now a barrier {for departing} from {the  {metastable} buckled configuration (of course, assuming that the buckled filament retains its integrity, which may not be the case in great many practical situations)}. Using the same principle as before this transition will occur at an appreciable rate only when the barrier is of the order of $k_B T$ in height. This occurs at a different force, which we call $f^{**}$, that is to say the system shows hysteresis {(as is typical for first-order phase transitions)}.

\begin{figure}
\centering
\resizebox{0.4\textwidth}{!}{\includegraphics{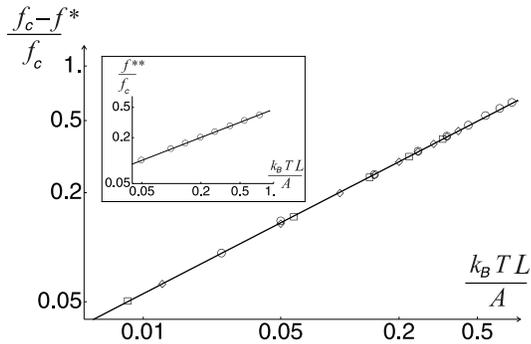}}  \caption{A plot of $1-f^*/f_c$, and inset $f^{**} / f_c$, against $k_B T L / A$ on log-log axes. The straight lines illustrate the simple scaling relations highlighted in the text do indeed hold.} \label{scaling}
\end{figure}

The central result of this work is that we find that both $f^*(T)$ and $f^{**}(T)$ follow the simple scaling laws of the form
\begin{equation}
\begin{array}{ccc}
f^* / f_c & = &  1 -  c \ (k_B T L / A)^{\nu}  \\
\\
f^{**} / f_c & =  &  d ( k_B T L /A )^{\eta}\\
\end{array}
\end{equation}
{These relations are illustrated} in Figure~\ref{scaling} where we plot $1-f^*/f_c$, and inset $f^{**} / f_c$, against $k_B T L / A$ {(which is the only non-dimensional combination of model parameters)} on log-log axes and obtain straight lines. The best fit values for the parameters are $c = 1.11$, $\nu = 0.56$, $d = 0.69$, $\eta = 0.50$. As expected, in the limit of a very stiff filament (large $A$)  $f^* \rightarrow f_c$ recovering the macroscopic behavior of {a purely elastic} rod under compression. How the critical buckling force $f^*$ is reduced at finite temperatures is the {unusual result}. We  {have to emphasize} that the exponent $\nu = 0.56$ is {truly universal and}  {remains} independent of what cutoff we choose for our ``all of nothing'' kinetics. Indeed, redefining the cutoff to be, say, $2 k_B T$ would simply rescale our units of temperature, changing the value of {the constant prefactor} $c$ alone.

For filaments with $A / k_B T L \approx 1$ the effects of non-zero {thermal fluctuations} become {increasingly} important. The effects described in this work should be experimentally observable in microscopic filaments such as microtubules, actin filaments and carbon nanotubes that are in this regime {of bending rigidity}. For such filaments this work suggests that the buckling force will no longer scale like $L^{-2}$ as it does for {the Euler buckling of} macroscopic rods, but like $L^{-2}(1-c(L/l_p)^{0.56})$ where $c$ is a constant of order unity {and $l_p=A / k_B T$ is the nominal persistence length of the filament}.

\section{Conclusions}

We began this paper aiming to examine buckling in filaments where thermal energy $k_B T$ becomes comparable to the energy scales of bending $A / L$. Within the {mean field} approximation of global inextensibility of such semiflexible chains we have found that there is a simple algebraic expression for the free energy of a filament subject to a constant force. This expression captures the correct physics of filaments in both the flexible limit ({flexible} Gaussian chains) and stiff limit (rigid {elastic} rods).

For any non-zero compression, the expression obtained for the free energy of single semiflexible chains develops a local minimum corresponding to a buckled state in addition to the minimum corresponding to the unbuckled state. The {buckling} transition is therefore determined by an escape from a local minimum in free energy to the global minimum. We have {adopted} a simple ``all-or-nothing'' kinetic model for this transition, {in which} the transition occurs if the free energy barrier between the states is less than $k_B T$. This suggests that what determines whether a filament buckles is kinetics: how quickly the filament can equilibrate into the global free energy minimum.

Based on this hypothesis, we find that the critical buckling force for filaments in a thermal environment is no longer the classical macroscopic expression $f_c = \pi^2 A / L^2$ obtained by Euler. Instead, the critical buckling force is lowered by a factor $ \sim (k_B T L / A)^{0.56}$ {with a universal scaling exponent}. This effect should be experimentally observable in semiflexible filaments and could have an importance for a wide class of filaments that are the subject of current research.


\end{document}